\documentclass[twocolumn,prb,showpacs,amsmath,amstex,amssymb,mathfonts,superscriptaddress]{revtex4-1}

\pdfoutput=1
\usepackage{amsthm,color,amsfonts,graphicx,verbatim}
\usepackage{amsmath}
\usepackage{amssymb}
\usepackage{amsthm}

\usepackage[normalem]{ulem}

\usepackage{amsfonts}
\usepackage{listings}
\usepackage{enumerate}
\usepackage{latexsym}
\usepackage{bm}
\usepackage{graphicx}
\usepackage[breaklinks,colorlinks = true,linkcolor = red,urlcolor=cyan,citecolor=red]{hyperref}
\newcommand{\beq}{\begin{equation}}
\newcommand{\eeq}{\end{equation}}
\newcommand{\beqn}{\begin{eqnarray}}
\newcommand{\eeqn}{\end{eqnarray}}

\newcommand{\Ket}[1]{\left|#1  \right>}

\newcommand{\Braket}[1]{\left<#1  \right>}

\newcommand{\<}{\langle}

\newcommand{\up}{\uparrow}
\newcommand{\down}{\downarrow}
\renewcommand{\>}{\rangle}
\renewcommand{\(}{\left(}
\renewcommand{\)}{\right)}
\renewcommand{\[}{\left[}
\renewcommand{\]}{\right]}

\newcommand{\Z}{\mathbb{Z}}
\newcommand{\T}{\mathcal{T}}

\newcommand{\C}{\mathcal{C}}
\newcommand{\SL}{\mathcal{S}}

\renewcommand{\paragraph}[1]{\vspace{4pt}\noindent {\it #1.---}}

\begin{document}
\title{Particle-hole symmetry, many-body localization, and topological edge modes}

\author {Romain Vasseur}
\affiliation{Department of Physics, University of California, Berkeley, CA 94720, USA}
\affiliation{Materials Science Division, Lawrence Berkeley National Laboratories, Berkeley, CA 94720}
\author{Aaron J. Friedman}
\affiliation{Department of Physics and Astronomy, University of California, Irvine, CA 92697, USA}
\author{S. A. Parameswaran}
\affiliation{Department of Physics and Astronomy, University of California, Irvine, CA 92697, USA}
\affiliation{California Institute for Quantum Emulation (CAIQuE), Elings Hall, University of California, Santa Barbara, CA 93106, USA}
\author{Andrew C. Potter}
\affiliation{Department of Physics, University of California, Berkeley, CA 94720, USA}

\date{\today}
\begin{abstract}
We study the excited states of interacting 
fermions in one dimension 
with particle-hole symmetric disorder (equivalently, random-bond XXZ chains) using a combination of renormalization group methods and exact diagonalization.
Absent interactions, the entire many-body spectrum exhibits  infinite-randomness quantum critical behavior with highly degenerate excited states. We show that though interactions are an irrelevant perturbation in the ground state, they drastically affect the structure of excited states:  even arbitrarily weak interactions split the degeneracies in favor of thermalization (weak disorder) or spontaneously broken particle-hole symmetry, driving the system into a many-body localized spin glass phase  (strong disorder). In both cases, the quantum critical properties of the non-interacting model are destroyed, either by thermal decoherence or spontaneous symmetry breaking. This system then has the interesting and counterintuitive property that edges of the many-body spectrum are \emph{less} localized than the center of the spectrum. 
We argue that our results rule out the existence of certain excited state symmetry-protected topological orders.
\end{abstract}
\maketitle

\section{Introduction}
Many-body localization (MBL) extends the concept of single particle (Anderson) localization due to random chemical potentials~\cite{PhysRev.109.1492} to the excited states of isolated interacting quantum systems~\cite{FleishmanAnderson,Gornyi,BAA}. MBL systems raise the compelling prospect of 
supporting quantum coherent information storage and processing~\cite{PhysRevLett.111.127201,PhysRevB.90.174202,PhysRevLett.113.147204,PhysRevB.91.140202,PhysRevX.4.041021,2015arXiv150806992C,2015arXiv150806995Y}, and nontrivial quantum order~\cite{HuseMBLQuantumOrder,BauerNayak,PekkerRSRGX,PhysRevB.89.144201,BahriMBLSPT,2015arXiv150505147S,2015arXiv150600592P} in highly excited states far from thermal equilibrium~\cite{2014arXiv1404.0686N}. Moreover, phase transitions between MBL states~\cite{PekkerRSRGX,PhysRevLett.112.217204,PhysRevLett.113.107204,QCGPRL} (or between MBL and thermalizing systems~\cite{PhysRevB.75.155111,PalHuse,MeanFieldMBLTransition,Luitz,VHA,PVPtransition}) represent new classes of non-equilibrium quantum critical behavior.

A natural generalization of random potential localization is particle-hole symmetric (PHS) disorder such as that due to random hopping amplitudes (or random vector potentials 
 in dimensions higher than one). In one dimension and in the absence of interactions, PHS disorder 
 does not fully localize single-particle states at zero energy, resulting in a marginally localized random-singlet phase with infinite randomness quantum critical properties~\cite{FisherRSRG2,PhysRevB.56.12970}. In this paper, we examine the fate of the highly excited states of this marginally localized phase~\cite{NandkishorePotterScaling} in the presence of interactions by studying an equivalent problem, the random-bond XXZ spin-$\frac{1}{2}$ chain 
\begin{equation} 
H=\sum_{i=1}^{L-1} J_i \left( S_i^x S_{i+1}^x+S_i^y S_{i+1}^y+ \Delta_i S_i^z S_{i+1}^z\right),
\label{eqXXZ}
\end{equation}
where $S_i^\mu = \frac{1}{2}{\sigma_i}^\mu$, and $\sigma_i^\mu$ with $\mu = x,y,z$ are the standard Pauli matrices. We consider open boundary conditions. In addition to spin conservation, Hamiltonian (\ref{eqXXZ}) has an Ising symmetry generated by ${\mathcal C}=\prod_i \sigma_i^x$.
A Jordan-Wigner transformation maps (\ref{eqXXZ}) into a spinless fermion chain with nearest-neighbor interactions (that vanish for $\Delta_i=0$), with  $\mathcal{C}$ now playing the role of PHS. 
In thermal equilibrium and at zero temperature, interactions are 
 an irrelevant perturbation and do not affect the ground state critical properties~\cite{FisherRSRG2}. However, in the absence of interactions, the excited states are highly degenerate due to the combination of single-particle integrability and symmetry, and hence even weak interactions can be expected to dramatically modify the dynamical properties of this system. 

Using a combination of real-space renormalization group (RSRG) arguments and exact diagonalization we show that arbitrarily weak interactions necessarily destroy the random-singlet critical properties in excited states, either by inducing 
thermalization (at weak disorder) or by spontaneously breaking 
PHS (strong disorder). 
In the latter case, this leads to a counterintuitive scenario wherein 
 the ground state is {\it less} localized (more entangled) than excited states. In addition, the ground-state random singlet phase can be thought of as a phase transition between a certain 1D symmetry protected topological insulator with chiral symmetry and a trivial insulator, and hence understanding its dynamical behavior will also shed light on questions of extending symmetry-protected topological (SPT) order (and related Floquet SPT orders)  to highly excited states in MBL systems. We argue that the 
 spontaneous symmetry breaking inherent at strong disorder presents a fundamental obstacle to achieving this goal.
 
These results should be contrasted with a prior study of XXZ chains~\cite{VoskAltmanPRL13} that used a related dynamical RSRG method to argue that the quantum critical behavior of the non-interacting ground state extends to highly excited states. 
However, as noted in~\cite{VoskAltmanPRL13}, these dynamical RSRG results apply only to the fine-tuned N\'eel initial state which artificially removes the excited-state degeneracies from the dynamically accessible Hilbert space. We expect that our results reflect the true dynamical properties of generic ({\it i.e.}, not fine-tuned) states.
  

%

\section{Strong disorder renormalization group}

\subsection{RSRG-X}
The $T=0$ low-energy physics of the antiferromagnetic XXZ spin chain~\eqref{eqXXZ} is well understood in terms of a real-space renormalization group (RSRG) approach valid at strong disorder~\cite{FisherRSRG2}. The key idea is to focus on the strongest bond of the chain $\Omega = J_i$. Assuming strong disorder, this bond is typically much larger than its neighbors, $\Omega \gg J_{R}, J_{L}$ ($J_{R/L} \equiv J_{i\pm1}$), so to leading order we can diagonalize this strong bond by forming a singlet between the spins $S_i$ and $S_{i+1}$. Quantum fluctuations then induce an effective XXZ coupling between the spins $S_L=S_{i-1}$ and $S_R=S_{i+2}$. Iterating this procedure, the effective disorder strength grows under renormalization so that RSRG becomes asymptotically exact -- {\it i.e.} gives exact results for universal quantities~\cite{FisherRSRG2}. This approach was recently generalized to construct many-body excited states of random spin chains by observing that at each 
 step, it is possible to project the strong bond onto an excited-state manifold~\cite{VoskAltmanPRL13,PekkerRSRGX}. The resulting excited-state RSRG (RSRG-X~\cite{PekkerRSRGX}) iteratively resolves smaller and smaller energy gaps $\Omega$ and allows one to construct, in principle, the full many-body spectrum. 
Assuming $\Delta \equiv \Delta_i \neq \pm1$ to avoid resonances, projecting onto the eigenstates $\Ket{\uparrow\downarrow} \pm \Ket{\downarrow \uparrow}$ of the strong bond preserves the XXZ form of the effective interaction between  $S_L$ and $S_R$ with parameters $J_{\rm eff}=J_L J_R/((1 \mp \Delta)\Omega)$ and $\Delta_{\rm eff}=\Delta_L \Delta_R (\Delta \mp 1)/2$, respectively. Another possibility would be to project onto the zero-energy states $\Ket{+} = \Ket{\uparrow\uparrow}$,  $\Ket{-} = \Ket{\downarrow\downarrow}$, where these two degenerate states can be interpreted as components of a new effective superspin $S_{\rm eff}$ with a different $U(1)$ charge $S_z=\pm 1$ than the original UV spins $\frac{1}{2}$. Spin conservation implies that $S_{\rm eff}$ cannot be flipped by a first order process like $S_{L,R}^+ S^-_{\rm eff} + {\rm h.c.}$
Keeping track of all the symmetry-allowed processes, we find the effective Hamiltonian
\begin{multline}
H_{\rm eff} = J_L \Delta_L S^z_L S^z_{\rm eff}+ J_R \Delta_R S^z_R S^z_{\rm eff}\\
+ \frac{J_L J_R}{\Omega (\Delta^2-1) }  \left[ \frac{S^+_L S^-_R}{2} + \Delta S^+_L S^+_R S^-_{\rm eff}+{\rm h.c.}\right]+ \dots
\label{eqHeff}
\end{multline}
where we have ignored second-order corrections to the Ising $S^z_{L,R} S^z_{\rm eff}$ terms. 

\begin{figure}[t!]
\includegraphics[width=1.0\columnwidth]{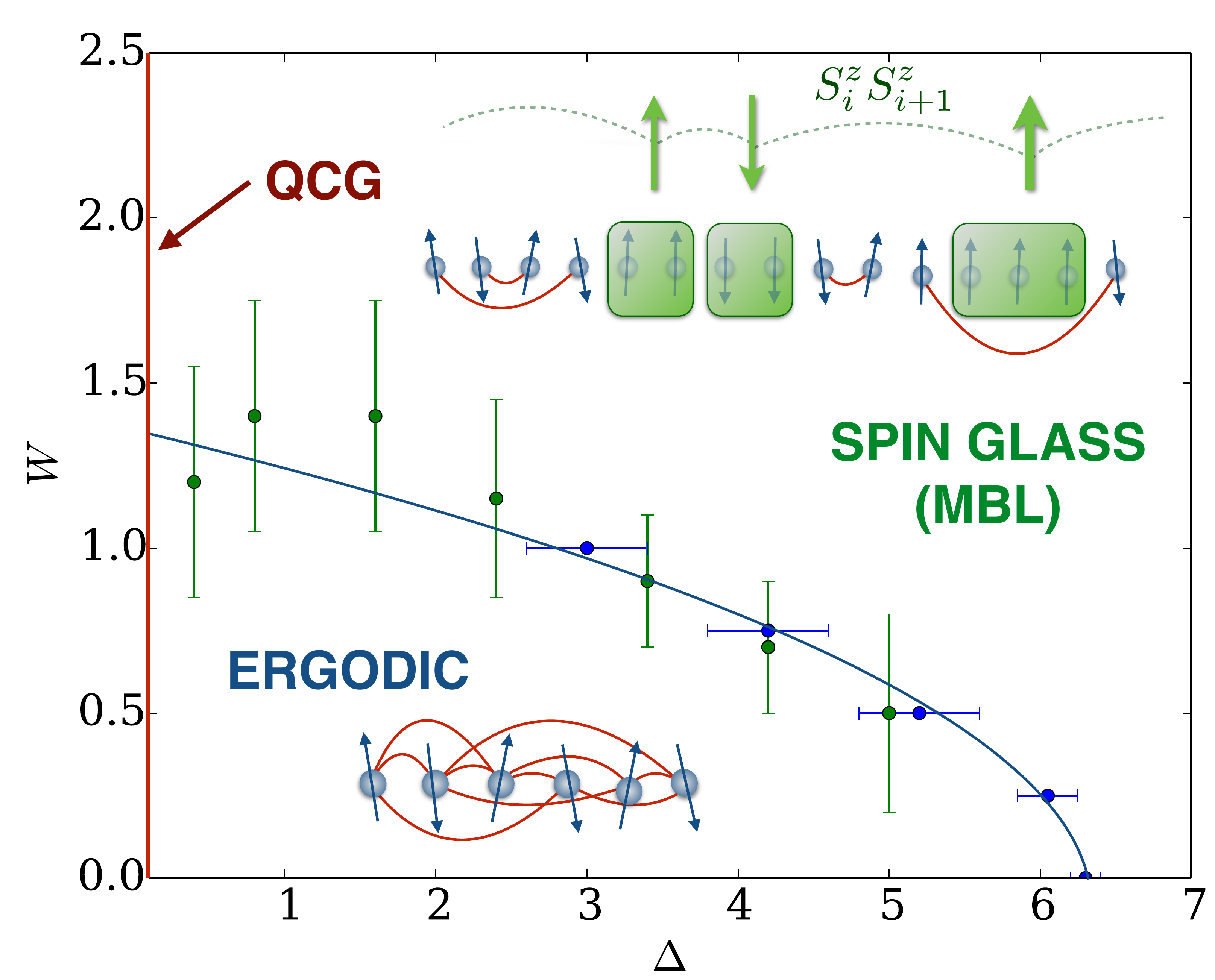}\vspace{-.1in}
\caption{\label{fig:PD} {\bf Phase diagram of the random-bond XXZ chain} at energy density $\epsilon=0.5$ from exact diagonalization results. The quantum critical behavior of the free model ($\Delta=0$) is destroyed by interactions, giving rise to either an ergodic phase at weak disorder where all spins are highly entangled, or to a many-body localized phase with spin glass order at strong disorder. {The phase boundary is estimated from finite-size crossings at constant $W$ (blue symbols) or constant $\Delta$ (green symbols).} The excited states in the spin glass phase consist of effective superspins (green spins) showing a random pattern of frozen magnetization varying from eigenstate to eigenstate.  }
\vspace{-.1in}
\end{figure}

In the non-interacting case~\cite{YichenJoel} ($\Delta_i=0$), the effective Hamiltonian always has the same XX form as the original one so that the procedure can be readily iterated.
 The sign of the $J$ coupling being essentially irrelevant, the flow equations for the couplings are identical to the groundstate ones.
This indicates that the random XX chain at finite energy density is a ``Quantum Critical Glass''~\cite{QCGPRL} (QCG), a critical variant of MBL with logarithmic scaling of the entanglement and power-law mean correlation functions. 
Crucially, the effective spins $S_{\rm eff}$ (created when projecting onto the $S_z=\pm 1$ excited states) completely decouple from the rest of the chain, thereby 
 producing an exponential degeneracy of the many-body eigenstates generated by RSRG-X.  This degeneracy is 
 a consequence of the PHS of the single-particle spectrum, 
 that dictates that single-particle energies come in pairs $(\epsilon,-\epsilon)$. The remainder of this paper focuses on investigating the fate of these extensive degeneracies upon the inclusion of interactions.

\subsection{Interaction-induced spin glass order}  From Eq.~\eqref{eqHeff}, we see that the interactions generate two new types of term: second-order couplings $S^+_L S^+_R S^-_{\rm eff}+{\rm h.c.}$ flipping the effective spin $S_{\rm eff}$ and more importantly Ising couplings  $S^z_{R,L} S^z_{\rm eff} $ generated at first-order in perturbation theory. To leading order, the effective Hamiltonian takes the form of a simple Ising coupling that will dominate over the 
much weaker second order flip-flop terms involving $S_L$, $S_R$.  
Although it is hard to keep track of all the multi-spin terms emerging after many RSRG-X iterations, the trend is already clear.
Namely, superspins made of $n>2$ aligned UV spins will be eventually generated in the course of the RG. Because of spin conservation, it is increasingly harder to flip these large superspins as this will involve higher-order processes in perturbation theory involving many super-spin clusters. This strongly suggests a physical picture of the excited states in terms of almost frozen superspins with strong Ising interactions, very weakly coupled by flip-flop terms generated at higher order in perturbation theory. The eigenstates would then consist of (super)spins showing a random pattern of frozen magnetization --- breaking the Ising symmetry --- varying from eigenstate to eigenstate. 

Such spontaneous breaking of PHS by interactions generates a random chemical potential term $\sum_i \mu_i S_i^z$  ({\it e.g.} in a mean field treatment $\mu_i = \sum_{j=i-1,i+1}J_{j}\Delta_{j}\<S^z_{j}\>$), which localizes the extended single-particle modes near zero energy and cuts off the quantum critical spin fluctuations at length scales longer than the spin-glass correlation length. Spontaneous PHS breaking appears to be the only route to an MBL phase in this model: in particular, single-spin terms $h_i S^{x,y,z}_i$ acting on the super-spins are forbidden by symmetry. This result implies that whereas the edges of the many-body spectrum are quantum critical with algebraic mean correlations, high energy density eigenstates are {\it more} localized, in sharp contrast with random-field MBL systems where higher energy densities tend to favor delocalization~\cite{BAA,Luitz}.



    \begin{figure}[t!]
\includegraphics[width=1.0\columnwidth]{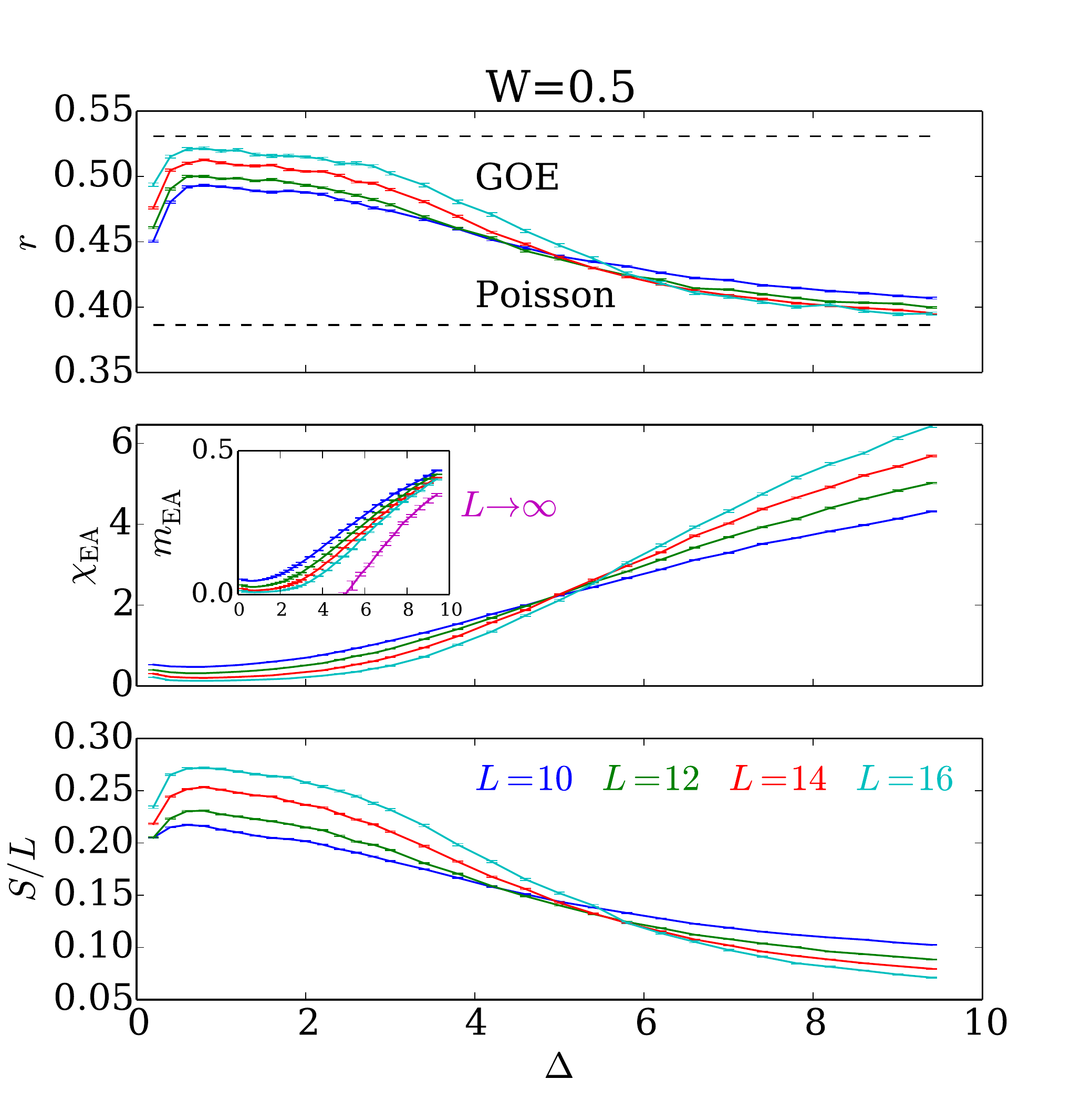}\vspace{-.1in}
\caption{\label{fig:Transition} {\bf Ergodic to spin glass (MBL) transition}. At weak disorder ($W=0.5$), our data are consistent with an ergodic to spin glass (MBL) transition as $\Delta$ is increased. {\it Top:} Ratio of consecutive level spacings showing a transition from GOE to Poisson statistics. {\it Middle:}  Scaling of $\chi_{\rm EA}$ showing a divergence with system size in the localized phase. Inset: Extrapolations of $m_{\rm EA}$ with $L^{-1}$ finite-size corrections (see text) are consistent with spin glass order in the MBL phase. {\it Bottom:} {Finite-size scaling of the entanglement entropy}.  }
\vspace{-.1in}
\end{figure}

\section{Numerical results}
Though the above argument based on RSRG-X strongly suggests that even infinitesimally weak interactions will destroy the quantum critical glass behavior of the random XX spin chain and lead to spin glass order instead, it is hard to explicitly track all the higher-order terms generated during the renormalization process that could (in principle) flip the super-spins.  In order to clarify this issue, we now turn to numerical exact diagonalization methods to study~\eqref{eqXXZ}. We draw the couplings $J_i \in (0,1]$ from the power-law distribution $P(J)=\frac{1}{W} \frac{1}{J^{1-1/W}}$ and we choose $\Delta_i$ to be uniformly distributed in the interval $[-\Delta,\Delta]$. We also restrict to even $L$ and $\sum_i S_i^z=0$, and consider the even sector of the $\mathbb{Z}_2$ symmetry ${\cal C}$.
 For each disorder realization, we first calculate the extremal energies $E_{\rm min}$ and $E_{\rm max}$ and define the normalized energy density $\epsilon =(E-E_{\rm min})/(E_{\rm max}-E_{\rm min})$. We then use the shift-invert method~\cite{Luitz} to obtain the 50 eigenstates with energy closest to $\epsilon=0.5$, corresponding to the middle of the many-body spectrum. Results are averaged over at least $2 \times 10^3$ disorder realizations.

    \begin{figure*}[t!]
\includegraphics[width=2.0\columnwidth]{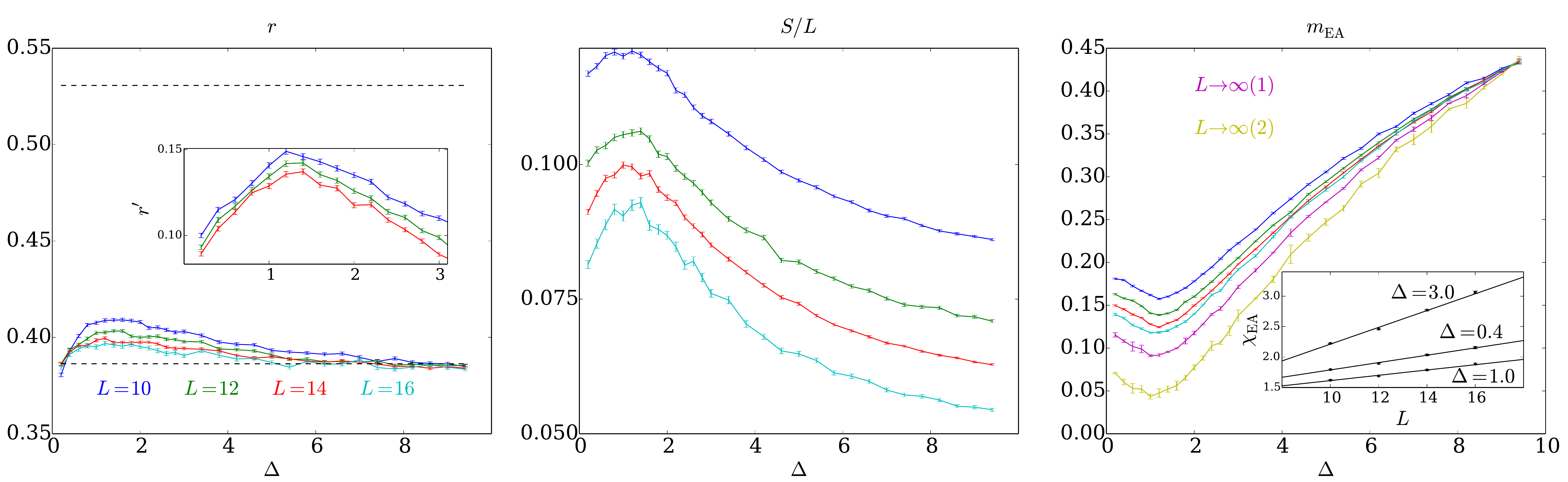}\vspace{-.1in}
\caption{\label{fig:SG} {\bf Strong disorder spin glass phase ($W=2.0$)}. 
 {\it Left}: Poisson statistics of the level spacings. Inset: when not restricted to a given ${\mathbb Z}_2$ sector, the gap ratio $r^\prime$ decreases with system size (well below the Poisson value), signaling pairing of the excited eigenstates.
 {\it Middle}: Sub-extensive scaling of the entanglement entropy. {\it Right}: Extrapolations of the spin glass order parameter $m_{\rm EA}=\chi_{\rm EA}/L$  are consistent with non-vanishing values in the thermodynamic limit for all values of $\Delta>0$, indicating spin glass order. Extrapolations are performed using $1/L^2$ (1) and $1/L$ (2) finite-size corrections. We note that the small dip around $\Delta \approx 1$ is naturally accounted for by the enhanced probability of local resonances $\Delta_i \approx 1$~\cite{QCGPRL} (see text).
 Inset: Linear scaling of $\chi_{\rm EA}$ with system size, consistent with spin glass order.}
\vspace{-.1in}
\end{figure*}

To distinguish between ergodic and non-ergodic phases we measure the level spacing parameter $r_{n}=\min(\delta_n,\delta_{n+1})/\max(\delta_n,\delta_{n+1})$~\cite{PhysRevB.75.155111} characterizing the ratio between consecutive level spacings $\delta_n = E_n -E_{n-1}$ averaged over energy levels $n$. Its disorder-averaged 
value changes from 
that characteristic of random matrices in the Gaussian orthogonal ensemble, $r_{\rm GOE}\simeq 0.5307$~\cite{PhysRevLett.110.084101} in the ergodic phase, to $r_{\rm Poisson}=2 \ln 2 -1 \simeq 0.3863$ (reflecting absence of level repulsion) in the 
MBL regime. We also compute the bipartite entanglement entropy $S_n= - {\rm tr} \rho_n \ln \rho_n $, where $\rho_n$ the reduced density matrix in the $n^{\rm th}$ eigenstate after tracing over half of the system. The entanglement scales as $S_n\sim 1,\log L$, and $L$ for MBL, QCG, and thermalizing systems respectively. To characterize 
the spin glass order, we introduce 
an Edwards-Anderson-like order parameter, 
%
\begin{equation}
m_{\rm EA}  = \frac{1}{L^2} \sum_n \sum_{i \neq j} \Braket{n \left| \sigma_i^z \sigma_j^z  \right|n}^2,
\label{eqEA}
\end{equation}
which tends to a constant (zero) in the thermodynamic limit for a spin-glass ordered (disordered) phase. 
 (We also consider the auxiliary quantity $\chi_\text{EA}\equiv Lm_\text{EA}$~\cite{PhysRevLett.113.107204}, which can in principle distinguish short-range spin glass order from certain types of quasi-long range order.)

\subsection{Phase diagram}

The results are summarized in the phase diagram of Fig.~\ref{fig:PD} (see Appendix~\ref{AppNumerics} for details). For weak disorder ($0\leq W\lesssim 1.5$), and $\Delta < \Delta_c(W)$, we find GOE level statistics, extensive entanglement, and vanishing spin glass order signaling a thermal phase. In this range of $W$,  increasing $\Delta$ drives an MBL transition to a spin glass-ordered phase at $\Delta=\Delta_c(W)$, heralded by a crossing in the finite size scaling plots of $r$, $S$, and $\chi_\text{EA}$ (Fig.~\ref{fig:Transition}). 

Interestingly, our numerics strongly suggest that for weak enough disorder, arbitrarily weak interactions lift the degeneracies of the non-interacting case and lead to thermalization. This is natural since at weak disorder and $\Delta=0$, degenerate PHS-conjugate pairs of orbitals that are either doubly occupied or both empty (corresponding to superspins in the RSRG-X language) have large spatial extent and overlap with many other degenerate pairs of orbitals. Then, upon the inclusion of interactions there are many strongly overlapping resonances that lead to thermalization. In other words, at weak disorder each orbital typically overlaps with many others, such that the higher-spin $S^+_LS^+_R \dots S^-_{\rm eff}$ type flip-flop terms are no longer strongly suppressed by many powers of a small parameter, and the massive degeneracy of the non-interacting case can be lifted by quantum fluctuations which naturally lead to thermalization. For strong disorder however, each degenerate pair of orbitals are sharply localized and interacts mainly with its nearest neighbors through predominantly Ising interactions leading to the spin glass MBL phase discussed above.  

\subsection{Strong disorder regime}

For strong disorder ($W\gtrsim 1.5$), we observe a clear finite-size scaling trend towards Poisson level statistics, and sub-extensive entanglement entropy. In this strong-disorder regime, our RSRG-X predictions should apply, and we therefore expect an MBL spin glass phase (Fig.~\ref{fig:SG}). 
  To distinguish between QCG and MBL phases, we examine the scaling of the spin glass order parameter with system size. Whereas for an MBL phase with long-range spin glass order $\lim_{L\rightarrow \infty} m_\text{EA} \neq 0$, for a QCG with only algebraic quasi long-range order, $m_\text{EA}\sim L^{-\alpha}$ ($\chi_\text{EA}\sim L^{1-\alpha}$). We observe that $\chi_{\text{EA}}$ 
 clearly grows with system size, inconsistent with a QCG with $\alpha>1$. In particular, this observation rules out a QCG in the same universality class as the random XX case ($\Delta=0$)~\cite{VoskAltmanPRL13}, which would have $\alpha=2$~\cite{FisherRSRG2}. 
 From our RSRG-X scenario, we expect two types of finite size corrections to $m_\text{EA}$: $1/L$ terms coming from short-range ordered regions 
 , and $1/L^2$ terms from the vestige of random-XX QCG.
 Extrapolating our data for $m_{\rm EA}$ using fits to either $1/L$ or $1/L^2$ finite-size corrections predicts a non-vanishing limiting value of $m_{\rm EA}$, suggesting spin-glass order for all $\Delta>0$  (Fig.~\ref{fig:SG}).  Though our data is perfectly consistent with linear growth of $\chi_\text{EA}=Lm_\text{EA}\sim L$, we cannot definitively rule out a more exotic QCG phase with $\alpha \ll 1$, distinct from the $\Delta=0$ XX random singlet phase. 
 
 

 In this spin glass (MBL) phase, the eigenstates for large systems should be cat states $\Ket{n}_\pm = (\Ket{n} \pm \C  \Ket{n})/\sqrt{2}$ that are even/odd under the ${\mathbb Z}_2$ symmetry generated by $\C = \prod_i \sigma_i^x$, where $\Ket{n}$ is some eigenstate-dependent pattern of $\sigma^z$ magnetization (with some background of random-singlet spins). The energy splitting between the two true eigenstates $\Ket{n}_\pm$ is exponentially small in system size and scales as $\sim {\rm e}^{-L/\xi}$ with $\xi$ the localization length, implying that the broken-symmetry state $\Ket{n}$ becomes metastable in the limit of large systems~\cite{HuseMBLQuantumOrder}. Meanwhile, the level spacing scales as $\delta \sim {\rm e}^{-(\ln 2) L }$ at ``infinite temperature'' (corresponding to our choice of normalized energy density $\epsilon=0.5$). At strong disorder, the localization length in the spin glass phase should be small and we therefore expect the eigenstates to be ``paired''~\cite{HuseMBLQuantumOrder}: the level spacing between each doublet is exponentially small  compared to the typical level spacing. This implies that the $r$ ratio should vanish, provided one does not restrict to a given ${\mathbb Z}_2$ sector (recall that up to now, we worked in the even sector of the particle-hole symmetry $\C $). In a quantum critical glass phase (with quasi-long range order), we expect the energy splitting of these quasi-doublets to become stretched-exponential~\cite{FisherRSRG1,PekkerRSRGX}, {\it i.e.} much larger than the many-body level spacing, thereby forbidding a regime with paired eigenstates. The $r$ ratio thus provides us with a clear way to distinguish true long-range spin glass order from the quasi-long range order of a quantum critical glass in small systems.
 
We checked that for sufficiently strong disorder ($W=2.0$ and $W=2.5$) where there is no sign of an ergodic phase, the $r$ ratio computed in the full spectrum of the $S_z=0$ sector (denoted by $r^\prime$) indeed decreases with system size, even for small values of $0.2 \leq \Delta \leq 1.0$ (Fig.~\ref{fig:SG}) where a quantum critical glass phase was previously predicted~\cite{VoskAltmanPRL13}. This strongly supports our claim of a spin glass phase extending all the way to infinitesimal $\Delta$.

We also remark that the small dip around $\Delta \approx 1$ in Fig.~\ref{fig:SG} is naturally explained by the enhanced probability of local resonances $\left| \Delta_i \right| \approx 1$. Recall that uniform anisotropies $\Delta_i=1$, $\forall i$ with global SU(2) symmetry are known to lead to thermalization for arbitrary disorder strength~\cite{VoskAltmanPRL13, PhysRevB.89.144201,QCGPRL}. In our case of inhomogeneous anisotropies distributed uniformly in $[-\Delta, \Delta]$, the probability of having a resonance $1-\epsilon<\left| \Delta_i \right | <1+\epsilon$  for small $\epsilon>0$ is obviously strictly zero if $\Delta<1-\epsilon$,  decays as $\sim 1/\Delta$ for $\Delta>1+\epsilon$, and therefore exhibits a maximum near $\Delta \approx 1$ (at $\Delta=1+\epsilon$).

\section{Constraints on protection of SPT order by MBL}

Our RSRG-X arguments and numerical 
results both show an inherent instability of the XX critical point towards a non-critical MBL spin glass upon the inclusion of interactions. Interestingly, these results imply a related instability of certain symmetry protected topological (SPT) orders, that one might have thought could emerge in 
highly excited states of MBL systems. Consider Eq.~\ref{eqXXZ}, with an even number of spins, dimerized hoppings $J_i = \frac{1}{2}J_i^{(0)}\(1+\delta_i(-1)^i\)$, and weak interactions ($\Delta_i\ll 1$). Then the ground state is topologically trivial for $\delta=\delta_i >0$, but exhibits SPT order with symmetry-protected spin-$\frac{1}{2}$ topological edge states for $\delta <0$ (see Appendix~\ref{AppSPT}). This model is dual to a 1D fermion SPT of class AIII
~ \cite{PerioTableKitaev,PerioTableSchnyder} via a standard Jordan-Wigner mapping, where the edge states are protected by the symmetry  $U(1) \times \Z_2^\mathcal{S}$ where $\mathcal{S}={\cal C} K$, with $K$ acting as complex conjugation (see Appendix~\ref{AppSPT}). 


In the perfectly dimerized limit, $\delta=-1$, the ground state consists of singlets on all dimerized bonds, with dangling spin-$\frac{1}{2}$ degrees of freedom at the left and right ends, and excitations are either non-degenerate $S^z=0$ triplets, $|\up_i\down_{i+1}\>+|\down_i\up_{i+1}\>$, or doubly degenerate $S^z=\pm 1$ triplets, $|\up_i\up_{i+1}\>,|\down_i\down_{i+1}\>$ on a strong bond. Moving away from the perfectly dimerized limit, $\delta\gtrsim -1$, these doubly degenerate $S^z=\pm 1$ bond-triplets weakly interact via virtual excitations of the intervening non-degenerate $S^z=0$ bonds. These interactions are strongly random, decaying exponentially in distance between the $S^z=\pm 1$ bonds, and symmetry dictates that these interactions be of XXZ form (plus less relevant multi-spin interactions). Thus the $S^z=\pm 1$ excitations form a new effective XXZ chain that, crucially, has no memory of the initial dimerization pattern $\delta_i$. According to the preceding sections of this paper, at finite energy density this effective XXZ chain will either thermalize (weak disorder) or spontaneously break symmetry (strong disorder); in both cases, the underlying SPT order is destroyed.

\section{Discussion}
 
We have argued that the notion of particle-hole symmetric Anderson localization does not extend to the MBL case. Even  though interactions are an irrelevant perturbation in the ground state, they drastically affect the structure of excited states leading either to thermalization at weak disorder or to spontaneously broken particle-hole symmetry at strong disorder, thereby destroying in both cases the quantum critical properties of the non-interacting model.

Our results also imply the instability of SPT order with $U(1) \times \Z_2^\mathcal{S}$ symmetry. 
Previous analyses of whether SPT order can extend to highly excited states of MBL systems~\cite{PhysRevB.89.144201,2015arXiv150505147S,2015arXiv150600592P} focused on whether it is possible to construct a locally integrable (commuting projector) ``fixed-point" model of the phase for which all excited states are localized with concurrent SPT order. Our present study furnishes an example where such a locally integrable model is possible (the perfectly dimerized state), but for which there are inherent degeneracies in the excitations that, upon weak perturbation away from the strictly integrable limit, result in spontaneous symmetry breaking. Our results also rule out the realization of certain stable Floquet SPT orders~\cite{PhysRevB.82.235114,PhysRevX.3.031005,2015arXiv150607647N,2015arXiv150803344K} with no equilibrium counterparts, such as those in driven systems, that require an MBL setting to avoid catastrophic heating~\cite{PhysRevLett.115.030402,PhysRevLett.114.140401,2014arXiv1412.4752A}. It would be very  interesting to investigate whether our results can be generalized to rule out the existence of PHS many-body localization and related excited state SPT orders in higher-dimensional systems. 

 \vspace{10pt}\noindent{\it Acknowledgements.---} We acknowledge helpful discussions with E. Altman, Y. Gannot, T. Morimoto, M. Serbyn and A. Vishwanath. This work was supported by 
 the Gordon and Betty Moore Foundation's EPiQS Initiative through Grant GBMF4307 (ACP), the DOE LDRD and Quantum Materials Programs at LBNL (RV), NSF Grant DMR-1455366 and the  President's Research Catalyst Award No. CA-15-327861 from the University
of California Office of the President (SAP).

\appendix

\section{Additional numerical results}

\subsection{Numerics and phase diagram}

\label{AppNumerics}
We computed the $r$ parameter, the entanglement entropy and the spin glass order parameter $m_{\rm EA}=\chi_{\rm EA}/L$ averaged over eigenstates at energy density $\epsilon=0.5$ and over disorder, for various values of the parameters $W$, $\Delta$ and $L$. The random couplings $J_i \in (0,1]$ are drawn from the power-law distribution $P(J)=\frac{1}{W} \frac{1}{J^{1-1/W}}$ and we choose $\Delta_i$ to be uniformly distributed in the interval $[-\Delta,\Delta]$. 

\begin{figure*}
\includegraphics[width = 6.5in]{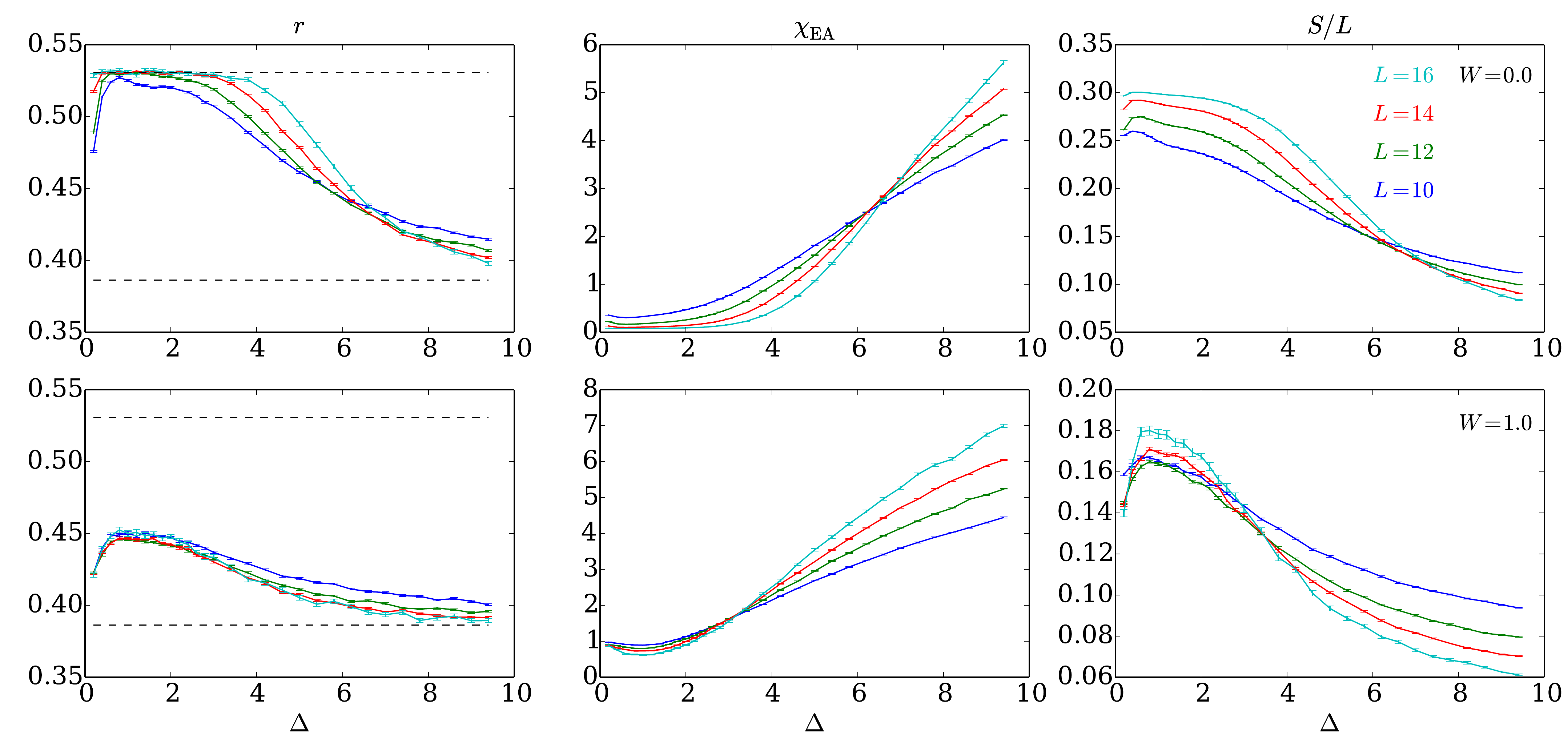}

\caption{Ergodic to spin glass (MBL) transition as a function of $\Delta$, for $W=0$ (top: uniform $J_i=1$) and $W=1$ (bottom). {\it Left:} Ratio of consecutive level spacings showing a transition from GOE to Poisson statistics. {\it Middle:}  Scaling of $\chi_{\rm EA}$ showing a divergence with system size in the localized phase.  {\it Right:} Finite-size entanglement crossover.}
\label{fig:sup1}
\end{figure*}

\begin{figure*}
\includegraphics[width = 3.5in]{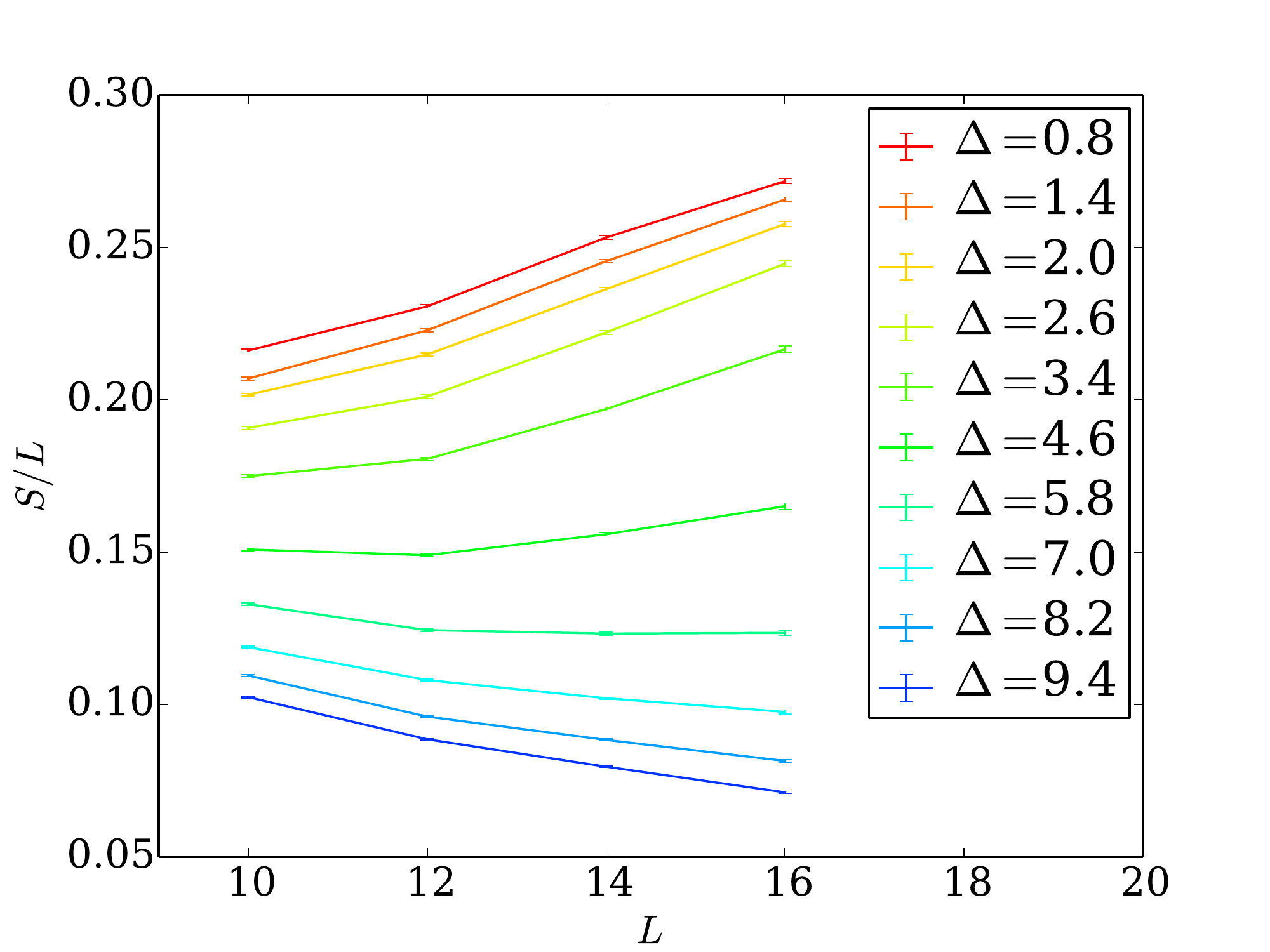}

\caption{Crossover between volume- and area-law scaling of the entanglement entropy as a function of $\Delta$ for $W=0.5$. }
\label{fig:supScaling}
\end{figure*}

\begin{figure*}
\includegraphics[width = 6.5in]{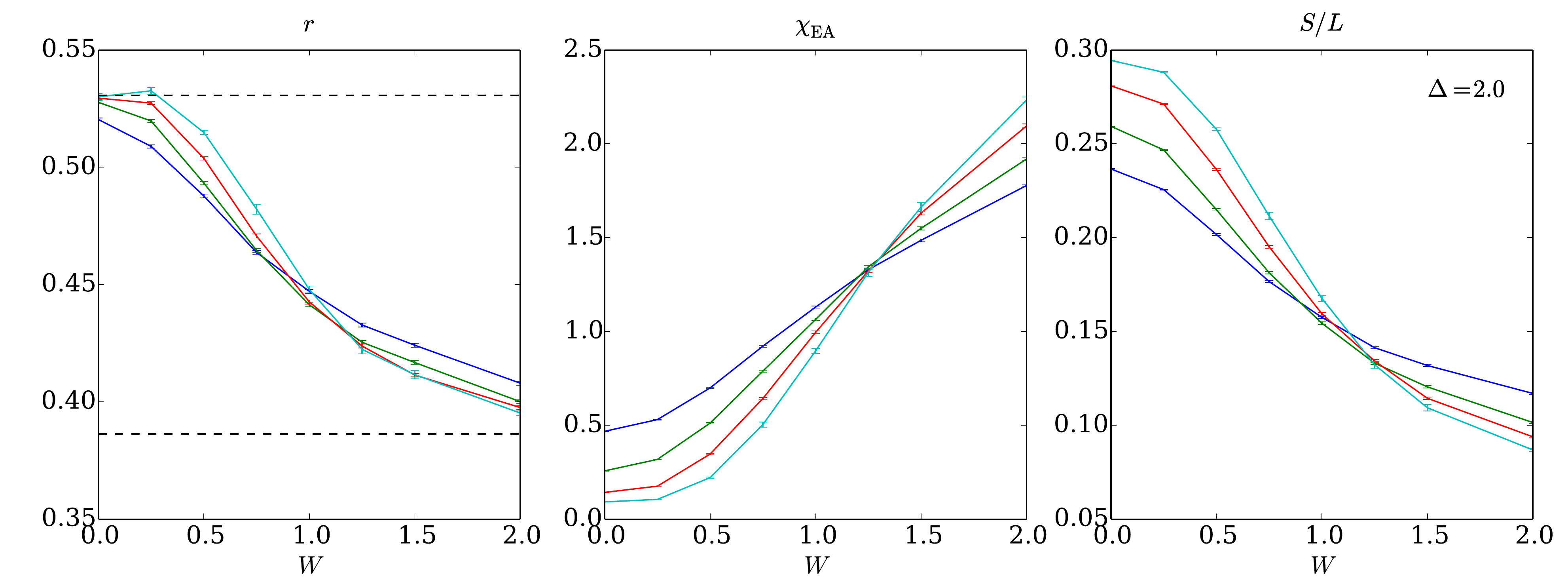}

\caption{Ergodic to spin glass (MBL) transition as a function of $W$ at fixed $\Delta=2.0$. }
\label{fig:sup2}
\end{figure*}

At weak enough value of the disorder strength ($ 0 \leq W \lessapprox 1.5$), we find evidence of an ergodic to MBL transition as a function of $\Delta$ (Fig.~\ref{fig:sup1}). For $\Delta$ smaller than a critical value $\Delta_c(W)$, the overall disorder strength is not enough to localize the system, and we observe clear signatures of a thermal phase (extensive entanglement entropy, GOE level statistics and $m_{\rm EA}=0$). For $\Delta > \Delta_c(W)$, the system is localized with Poisson level statistics, sub-extensive entanglement entropy indicating a breaking of ergodicity and diverging spin glass parameter $\chi_{\rm EA}$. The entanglement entropy as a function of $L$ shows a clear crossover from volume-law to sub-extensive behavior as a function of $\Delta$ (Fig.~\ref{fig:supScaling}).
This MBL transition can also be observed at fixed $\Delta$ by tuning $W$ (Fig.~\ref{fig:sup2}).

\subsection{Uniform anisotropies $\Delta_i=\Delta$}
\begin{figure*}
\includegraphics[width = 6.5in]{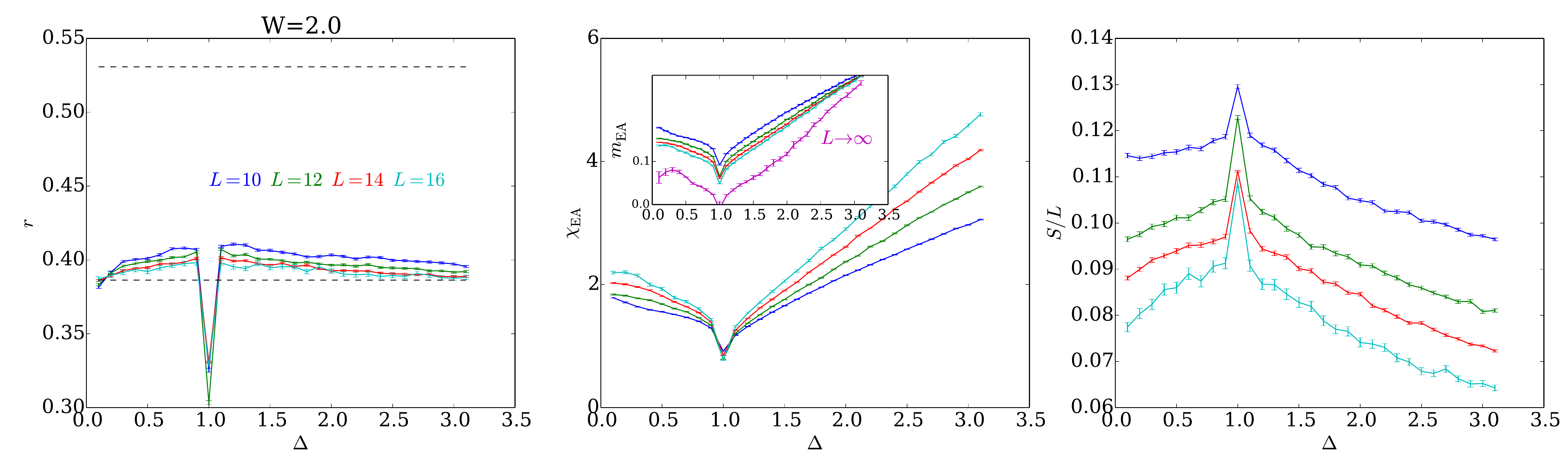}

\caption{Strong disorder regime ($W=2.0$) with uniform anisotropy $\Delta_i=\Delta$. Away from the pathological $SU(2)$-symmetric point $\Delta=1$, the results are qualitatively similar to the random $\Delta_i$ case, consistent with a many-body localized phase with spontaneously broken particle-hole symmetry at all values of $\Delta \neq 1$.}
\label{fig:supPureDelta}
\end{figure*}

Note that we took the anisotropy parameters  $\Delta_i$ to be random since randomness in $J_i$ generates randomness in $\Delta_i$ upon renormalization, so that we expect qualitatively similar conclusions for uniform $\Delta_i=\Delta$, except around the pathological SU(2) symmetric point $\left|\Delta\right|=1$ that is known to lead to thermalization for arbitrary disorder strength~\cite{QCGPRL} (see also~\cite{VoskAltmanPRL13, PhysRevB.89.144201}). To verify this numerically, we also considered the case of uniform anisotropies $\Delta_i=\Delta$ in the strong  randomness regime $W=2$ (Fig.~\ref{fig:supPureDelta}). We find that, away from the SU(2)-symmetric point $\Delta=1$, the results are qualitatively similar to the random $\Delta_i$ case as expected, with a spin-glass MBL phase at all values of $\Delta \neq 1$. Precisely at the SU(2)-symmetric point $\Delta=1$, the results are more intricate. Because of the different spin sectors $S^2=j(j+1)$ that do not mix with each others, it is natural to expect Poisson statistics in the $S^z=0$ sector even though the system should be thermalizing. We find numerically a $r$ ratio below the Poisson value, suggesting a finite-size segmentation of the spectrum. This is consistent with the fact that we observe sub-extensive entanglement for $\Delta=1$, which suggests that the system has not yet reached the scaling regime which should be dominated by almost classical large superspins. We nevertheless observe that the point $\Delta=1$ is much less localized than $\Delta \neq 1$ for the same disorder strength $W=2$. Moreover, our numerical results indicate the absence of spin glass order precisely at $\Delta=1$, also consistent with thermalization. We have also checked that taking $\Delta_i=\Delta$ does not modify the phase diagram shown in Fig.~\ref{fig:PD} quantitatively provided $\Delta \neq 1$. We leave a detailed numerical analysis of this interesting $SU(2)$-symmetric point for future work, and restrict ourselves to random anisotropies $\Delta_i $ which lead to qualitatively similar results but has the strong advantage of avoiding pathological features that are not the subject of interest for our study.  

\section{Fermion description and symmetry protected topological phases}
\label{AppSPT}

To be self-contained and make contact with our discussion in the main text, we briefly review the equivalent descriptions of the XXZ chain in terms of spinless fermions, and its connection to symmetry protected topological phases (SPTs) for the case of dimerized couplings.

The XXZ spin chain (Eq. 1 of main text) maps to an interacting fermion chain, via the standard Jordan-Wigner mapping $S^z_i = c^\dagger_ic_i-\frac{1}{2}$, $S^+_i = S^x_i+iS^y_i = \(\prod_{j<i}\sigma^z_j\)c_i^\dagger$:
\begin{align}
H_\text{spin} &= \sum_{i=1}^{2N-1} J_i \(S^x_iS^x_{i+1}+S^y_{i}S^y_{i+1}+\Delta_iS^z_iS^z_{i+1}\),
\nonumber\\
H_\text{fermion} &= \sum_{i=1}^{2N-1}J_i\[\frac{1}{2}\(c^\dagger_{i+1}c_i+ {\rm h.c.}\) \right. \notag
\\ & \left. +\Delta_i \(n_i-\frac{1}{2}\)\(n_{i+1}-\frac{1}{2}\)\],
\label{eq:Hform}
\end{align}
where we take $L=2N$ even. 
\subsection{Symmetries} 
We now discuss the symmetries of the above model. We will first describe the symmetries in the spin language, and then use the Jordan-Wigner mapping to obtain their action  on the fermion operators.

First, the spin chain has a $U(1)$ symmetry generated by $S^z$ rotations  $U_\phi=\prod_j e^{-i\phi \sigma^z_j/2}$, that corresponds to the conserved $z$-axis magnetization. The model also has $\Z_2$ time-reversal symmetry, implemented by $\T = K$, where $K$ is the antilinear operator representing complex conjugation, and an Ising ($\Z_2$) symmetry generated by $\C =\prod_j \sigma^x_j$. Note that the action of this symmetry flips the axis of the conserved spin, $\C  U_\phi \C ^\dagger = U_{-\phi}$.

Turning to the fermions, we see that the symmetries act on the second-quantized fermion operators as follows:
\begin{align}
U_\phi c_j U_\phi^\dagger &= e^{-i\phi}c_j \nonumber,\\
\T c_j\T ^{-1} &= c_j \,\,\,\,\,\text{with} \,\,\,\,\, \T i \T^{-1} = -i, \\
\C c_j\C ^{-1} &= (-1)^{j+1} c_j^\dagger.
\end{align}
Note that the time-reversal symmetry is anti-unitary, while $U_\phi$, $\C$ are unitary. In the fermionic language, $U(1)$ corresponds to the particle number conservation, while $\C$ corresponds to particle-hole symmetry.
In addition, we can construct an anti-unitary symmetry $\SL \equiv \C\times\T$, usually termed ``chiral'' or ``sublattice'' symmetry,
\begin{align}
\SL c_j\SL ^{-1} &= (-1)^{j+1} c_j^\dagger   \,\,\,\,\,\text{with} \,\,\,\,\, \SL i \SL^{-1} = -i.
\end{align}
Note that $\T^2=\C^2=1$ when acting on fermion operators, but that $\SL^2$ has no well defined action on $c_j$, because we can redefine $\SL^2$ by an arbitrary phase $e^{i\alpha}$ by combining it with a $U(1)$ rotation $\SL\rightarrow\tilde\SL=e^{i\alpha/2 \sum_j n_j}\SL$.\\

\noindent{\bf A note on nomenclature.}  We caution that there is a potentially confusing alternative terminology frequently used for non-interacting fermion systems, wherein the particle-hole symmetry $\C$ is called `antiunitary' while $\SL$ is called `unitary'. The alternative convention arises because the traditional symmetry classification of free fermion systems considers the action of symmetries on the first-quantized Hamiltonian 
$\mathcal{H}$, where the non-interacting second-quantized Hamiltonian $H$ is defined via $H = \sum_{i,j} c^\dagger_i \mathcal{H}_{ij} c_j$  (see also footnote on p.7 of \cite{TenfoldWay}). Given a unitary symmetry, $\mathcal{C}$, that interchanges particles and holes, $\C c_i\C^{-1}= \sum_j (U_\C^*)_{ij} c^\dagger_j$, then $\mathcal{H}$ satisfies $U_\C^\dagger \mathcal{H}^* U_\C = -\mathcal{H}$.

Owing to the complex conjugation on the LHS of the preceding expression, 
 the unitary symmetry $\C$ is sometimes termed `anti-unitary' in this context. Similarly, the anti-unitary symmetry $\SL$ implies  $U_\SL^\dagger \mathcal{H} U_\SL = -\mathcal{H}$, hence $\SL$ `looks unitary' when acting on $\mathcal{H}$. When referring to operators as unitary or anti-unitary, we will {\it always} refer to the action on the second-quantized operators, which is more appropriate for the generic case of interacting systems. Hence we will refer to $\C$ as unitary, and $\SL, \T$ as anti-unitary.

\subsection{Ground-state SPT Order}
When the hoppings in (\ref{eq:Hform}) are dimerized, $J_i \rightarrow \frac{J}{2}\(1-\delta(-1)^i\)$ with $1>\delta>0$,  the ground-state realizes an SPT phase with symmetry protected topological (complex) fermion zero mode end states. The non-trivial edge structure is most easily seen by considering the limit of zero interactions ($\Delta_i=0$) and strong dimerization, $\delta=1$. Here, the fermion Hamiltonian possesses a strictly localized complex fermionic zero mode $c_1$ ($c_{2N}$) on the left (right) side of the chain respectively. Focusing just on the left side of the chain, the fermionic zero mode $c_1$ spans a degenerate two-state Hilbert space $\{|\pm\>\}$ with $c_1|-\>=0$ and $|+\> = c_1^\dagger|-\>$. 

It turns out that the protection of these edge states relies only on $U(1)$ and the $\Z_2^\SL$ subgroup of $\Z_2^\C\times \Z_2^\T$, in the sense that we may break $\T$ and $\C$ separately so long as their product $\SL$ remains a good symmetry. Then, the relevant symmetry group is $U(1)\times \Z_2^\SL$, corresponding to class AIII in the Cartan notation
\cite{TenfoldWay,PerioTableKitaev,PerioTableSchnyder}. A classic example of a problem in this symmetry class is the Su-Schrieffer-Heeger model~\cite{PhysRevLett.42.1698}. %
In $d=1$, free fermion problems in this class have a $\Z$ classification, which reduces to  a $\Z_4$ classification upon including interactions~\cite{PhysRevB.81.134509}.

This zero energy edge mode has a projective implementation of the $U(1)\times  \Z_2^\SL$ symmetry group, which protects it from being gapped by any interaction with local bulk degrees of freedom, which all transform non-projectively and hence cannot couple in a symmetric fashion with the edge state.  The projective action of symmetry on the edge-state can be seen by considering just the $\Z_2$ subgroup of the $U(1)$ generated by the fermion number parity: $\mathcal{P}_F=e^{i\pi \sum_in_i}$. For the full system (and for any set of bulk degrees of freedom), the fermion parity operator commutes with charge conjugation, $\[\mathcal{P}_F,{\cal \SL}\]=0$. However, acting within the low-energy subspace spanned by the zero-mode states $|\pm\>$ of one end of the chain, we see that $\pm$ have opposite eigenvalue of $\mathcal{P}_F$: $\<+|\mathcal{P}_F|+\> = -\<-|\mathcal{P}_F|-\>$. On the other hand, ${\cal S}$ changes $c_1\rightarrow c_1^\dagger$, and hence exchanges $\SL|\pm\> = |\mp\>$. Hence, the symmetry group is implemented projectively at the end of the chain: $\mathcal{P}_F \SL \mathcal{P}_F \SL |\pm\> = (-1)|\pm\>$. This projective action of symmetry indicates that the zero-modes are topologically stable to any symmetry-respecting perturbation that does not close the bulk gap~\cite{PhysRevB.81.134509,PhysRevB.83.075102}. In particular, this phase and topological edge states exist in the ground state over a finite range of parameters near the perfectly dimerized limit (though the zero modes are generically only exponentially well localized to the edge).

The  phase described above is the elementary, $n=1$, ``root" phase of the 1D AIII chains. In the absence of interactions, we may combine any integer number $n$ of these phases to obtain a new non-trivial phase. For $n=2$ chains, let us denote by $c_{1}^{\dagger}$ and $d_{1}^{\dagger}$ the fermionic edge modes acting on the left side of the two perfectly-dimerized chains $n=1$ and $n=2$: the groundstate Hilbert space can then be written as $\{ |00\>, |10\>=c_1^\dagger  |00\>,  |01\>=d_1^\dagger  |00\>, |11\>=c_1^\dagger d_1^\dagger  |00\> \}$. Since $\SL c_1\SL ^{-1}=c^\dagger_1$ and $\SL d_1\SL ^{-1}=d^\dagger_1$, the sublattice symmetry ${\cal S}$ acts on the zero-mode states as $\SL|00\>= |11\>$, $\SL|11\>= -|00\>$, $\SL|10\>= |01\>$, and $\SL|01\>= -|10\>$. The $n=2$ phase therefore has $\SL^2=-1$ when acting on the edge states, whereas $\SL^2=1$ acting on any local bulk degree of freedom. Similarly, $n=3$ has a combination of $\{\mathcal{P}_F,\SL\}=0$ and $\SL^2=-1$ on the edge states. However, there is no projective action of symmetry for phases with $n=0\mod 4$, and hence in the presence of interactions these phases become equivalent to topologically trivial ones~\cite{PhysRevB.81.134509}. Our argument that the excited states of the $n=1$ phase with strong randomness are unstable to spontaneous symmetry breaking also applies to the other members ($n=2,3$) of this AIII SPT family, and rules out the protection of SPT order.



\bibliography{MBL}
\end{document}